\documentstyle[epsf]{article}

\newcommand{\Ell}{E_\parallel}      
\newcommand{\rhoGJ}{\rho_{{\rm GJ}}}  
\newcommand{\rlc}{\varpi_{\rm LC}} 
\newcommand{\inc}{\alpha_{\rm i}}  

\newcommand{\rhowSQR}{\rho_{\rm w}^2}
\begin{document}



\title{High-energy Emission from Pulsar Magnetospheres}

\author{KOUICHI HIROTANI\footnote{
            Postal address: 
            TIARA, Department of Physics, National Tsing Hua University,
            101, Sec. 2, Kuang Fu Rd., Hsinchu, Taiwan 300}\\
        ASIAA/National Tsing Hua University - TIARA,
        PO Box 23-141, Taipei, Taiwan\\
        {\it hirotani@tiara.sinica.edu.tw}
       }

\date{}
\maketitle

\vspace*{-0.5truecm}
\centerline{Modern Physics Letters A}
\centerline{Received: 01 May 2006}
\centerline{Accepted: 09 May 2006}
\bigskip

\begin{abstract}
A synthesis of the present knowledge on gamma-ray emission
from the magnetosphere of a rapidly rotating neutron star
is presented, 
focusing on the electrodynamics of particle accelerators.
The combined curvature, synchrotron, and inverse-Compton emission
from ultra-relativistic positrons and electrons, 
which are created by two-photon and/or one-photon pair creation 
processes, or
emitted from the neutron-star surface,
provide us with essential information on the properties of
the accelerator
--- electric potential drop along the magnetic field lines.
A new accelerator model,
which is a mixture of traditional inner-gap and outer gap models,
is also proposed,
by solving the Poisson equation for the electrostatic potential
together with the Boltzmann equations for particles and gamma-rays
in the two-dimensional configuration and 
two-dimensional momentum spaces.

\end{abstract}


\section{Pulsar nature and characteristics}
\label{sec:nature}
%
When a star has exhausted its nuclear fuel,
the interior thermal pressure can no longer support its own weight.
If the star is not very massive,
it collapses to the point at which electron degeneracy pressure 
halts gravitational collapse, and a white dwarf is formed.
However, beyond the critical mass~\cite{Chandra31},
even the electron degeneracy pressure becomes insufficient to 
support the star.
In the collapsing stellar core,
most of the electrons are absorbed by protons via inverse $\beta$ decay
and eventually
the neutron degeneracy pressure halts the gravitational 
collapse~\cite{Landau32,Baade34,Oppen39}.
For a progenitor star to collapse directly to a neutron star, 
its mass should be more than $10$ times greater than our Sun.
A neutron star, which can rotate with a period as fast as
$P \sim 1$~ms, is considered to be the only candidate for the
source of remarkably stable radio pulses, 
of which fastest period is 1.55~ms 
(PSR~J1939+2134~\cite{Lyne87}).
When it was determined that the period of the Crab pulsar
was slowly increasing~\cite{Richards69},
the identification of pulsars as rapidly rotating neutron stars was
essentially confirmed~\cite{Gold68,Gold69}.

Having established pulsars as rotating neutron stars,
the next thing to discuss is the pulsar characteristics deduced from
observations.
Since all known isolated pulsars show gradual spin down,
there must be some mechanisms causing the neutron star to lose their
rotational energy.
The rotational energy-loss rate is given by
\begin{equation}
  \dot{E}=I\Omega\dot\Omega=-(2\pi)^2I\dot{P}/P^3,
  \label{eq:Edot}
\end{equation}
where the neutron star is rotating with angular frequency 
$\Omega=2\pi/P$ and moment of inertia $I$.
For typical high-energy pulsars,
$P\sim 0.1$~s, $\dot{P}\sim 10^{-13}\,\mbox{s s}^{-1}$,
$I\sim 10^{45} \mbox{g\,cm}^2$; 
thus, we obtain $\dot{E}=-4\times 10^{36}\,\mbox{ergs s}^{-1}$,
which is enough to explain the observed $\gamma$-ray luminosities, 
$10^{32.5}\,\mbox{ergs s}^{-1} < L_\gamma < 10^{35}\,\mbox{ergs s}^{-1}$.
For an object as small as $r_\ast \sim 10$~km in radius 
to experience such a large torque,
it must have a strong coupling to their surroundings,
most likely through magnetic fields intrinsic to the neutron star.
If the neutron star is losing its rotational energy via
magnetic dipole radiation, 
the spin-down luminosity is given by
\begin{equation}
  L_{\rm dip}=k\Omega^4\mu^2/c^3,
  \label{eq:Ldip}
\end{equation}
where $k=2\sin^2\inc/3$ is commonly used 
for a vacuum rotating magnetic dipole with inclination $\inc$
with respect to the rotation axis.
Equating $L_{\rm dip}$ with $-\dot{E}$, we can infer the
magnetic moment, $\mu$, of the star~\cite{Ostr69}.
For $P=0.1$~s, $\dot{P}=10^{-13}\,\mbox{s s}^{-1}$,
$I=10^{45} \mbox{g\,cm}^2$, we obtain
$\mu = 2.6\times 10^{30}\,\mbox{G cm}^3$ and
the surface magnetic field strength, 
$B_\ast = 5.2\times 10^{12} (r_\ast/10\mbox{km})^{-3}$~G,
as the Newtonian value at the magnetic pole.
In what follows, we consider possible emission models of
isolated, rotating neutron stars,
after briefly mentioning the high-energy observations.

\section{Gamma-ray observations and physical processes}
\label{sec:obs}
\begin{figure} 
  \centerline{ \epsfxsize=8cm \epsfbox[0 50 600 600]{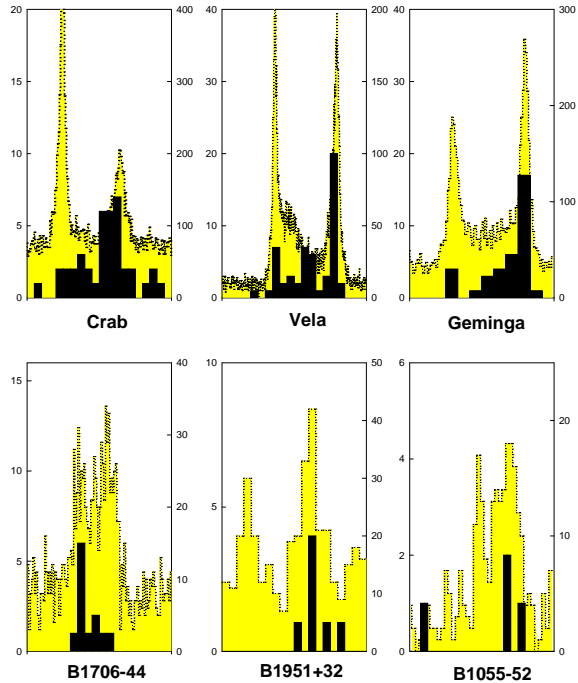}} 
\caption{
Light curves of six $\gamma$-ray pulsars 
above 100 MeV (dotted lines, right-hand scale) and 
above   5 GeV (dark histogram, left-hand scale).
Each panel shows one full rotation of the neutron star.
From Thompson (2004).  
\label{fig:lcurvs}
}
\end{figure}

The Energetic Gamma Ray Experiment Telescope 
and the imaging Compton Telescope 
aboard the Compton Gamma Ray Observatory 
have detected pulsed signals from at least seven rotation-powered pulsars
(Crab pulsar~\cite{Nolan93,Fierro98};
 PSR~B1509-58~\cite{Kuiper01};
 Vela pulsar~\cite{Kanbach94,Fierro98};
 PSR~B1706-44~\cite{Thompson96}; 
 PSR~B1951+32~\cite{Ramana95};
 Geminga pulsar~\cite{Mayer94,Fierro98};
 PSR~B1055-52~\cite{Thompson99}).
Four of them (Crab, Vela, Geminga, and B1951+32)
exhibit double-peaked light curves\cite{Kanbach99,Thompson01,Thompson04}
(fig.~\ref{fig:lcurvs}).
Since interpreting $\gamma$-rays should be less ambiguous
compared with reprocessed, non-thermal X-rays,
the $\gamma$-ray pulsations observed from these objects
are particularly important as a direct signature of 
basic non-thermal processes in pulsar magnetospheres.

Spectral energy distributions (SEDs) offers the key to an understanding
of the radiation processes.
Ref.~\cite{Thompson99} compiled a useful set of broad-band SED
for the seven $\gamma$-ray pulsars.
The most striking feature of these $\nu F_\nu$ plots 
is the flux peak above 0.1~GeV with a turnover at several GeV.
Various models~\cite{DH96,Romani96,ZC97,Higgins97,Higgins98,HS99}
conclude that these photons are emitted by the 
electrons or positrons accelerated above 5~TeV via curvature process.
Such ultra-relativistic particles could also cause 
inverse-Compton (IC) scatterings.
In particular, for young pulsars, their strong thermal X-rays
emitted from the cooling neutron-star 
surface~\cite{Becker97,Pavlov01,Kaminker02}
efficiently illuminate the gap to be 
upscattered into several TeV energies.
If their magnetospheric infrared (0.1--0.01~eV) 
photon field is not too strong,
a pulsed IC component could be unabsorbed to be detected 
with future ground-based or space telescopes.

The curvature-emitted photons have the typical energy of 
a few GeV.
Close to the star (typically within a few stellar radii),
such $\gamma$-rays can be absorbed by
the strong magnetic field ($> 10^{12}$~G) to materialize as a pair.
On the other hand, outside of this strong-field region,
pairs are produced only by the photon-photon collisions
(e.g., between the curvature-GeV photons and surface- or magnetospheric-
 keV photons).
The replenished charges partially screen the original 
acceleration field, $\Ell$.
If the created particles pile up at the boundaries of the potential gap,
they will quench the gap eventually.
Nevertheless, if the created particles continue to migrate
outside of the gap as a part of the global flows of charged particles,
a steady charge-deficient region will be maintained.
This is the basic idea of a particle acceleration zone 
in a pulsar magnetosphere.

\section{Representative emission models}
\label{sec:models}
\begin{figure} 
  \centerline{ {\epsfxsize=5.0truecm \epsfbox[0 0 420 200]{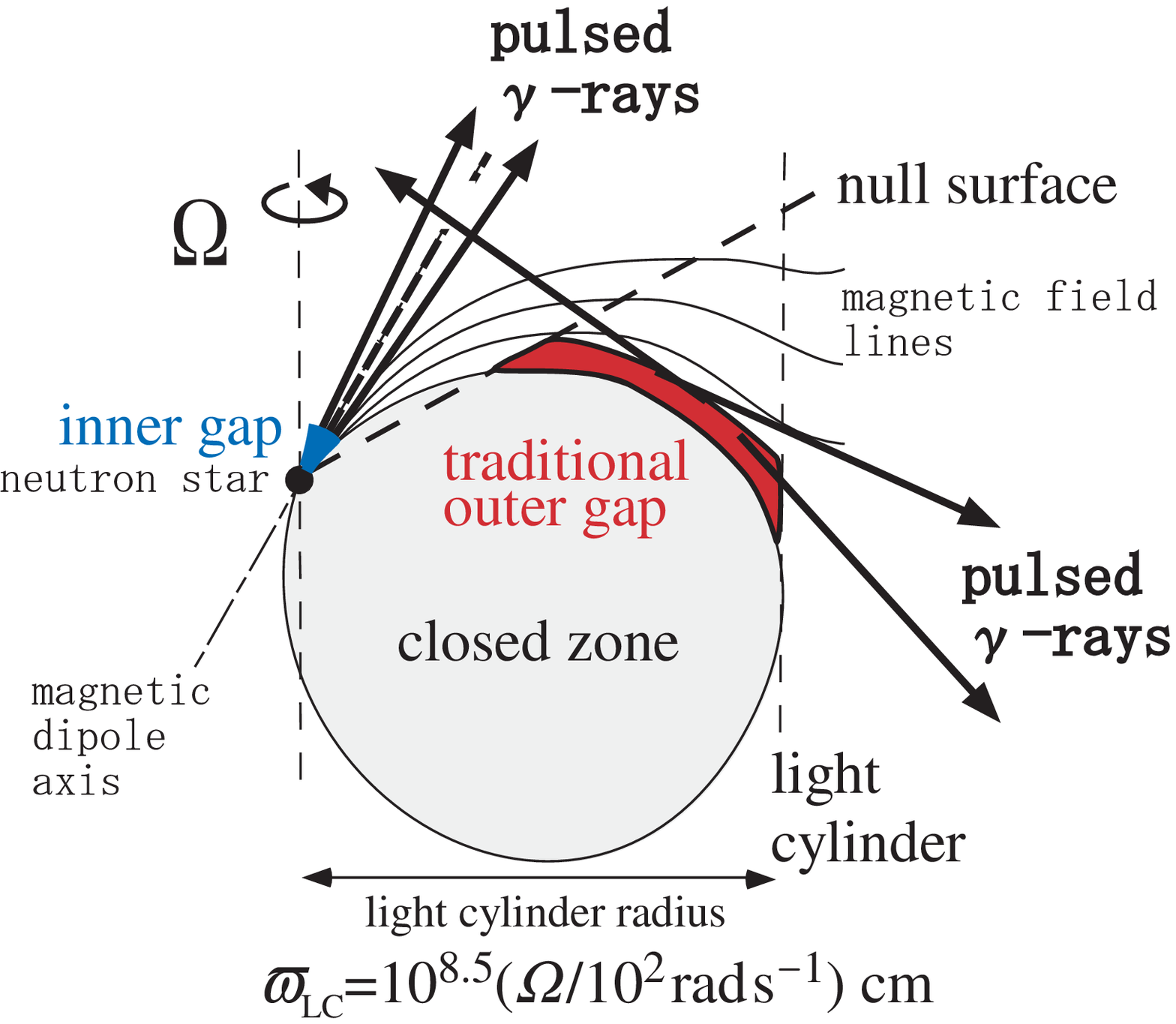}}
               \hspace*{0.5 truecm}
               {\epsfxsize=5.0truecm \epsfbox[0 0 420 400]{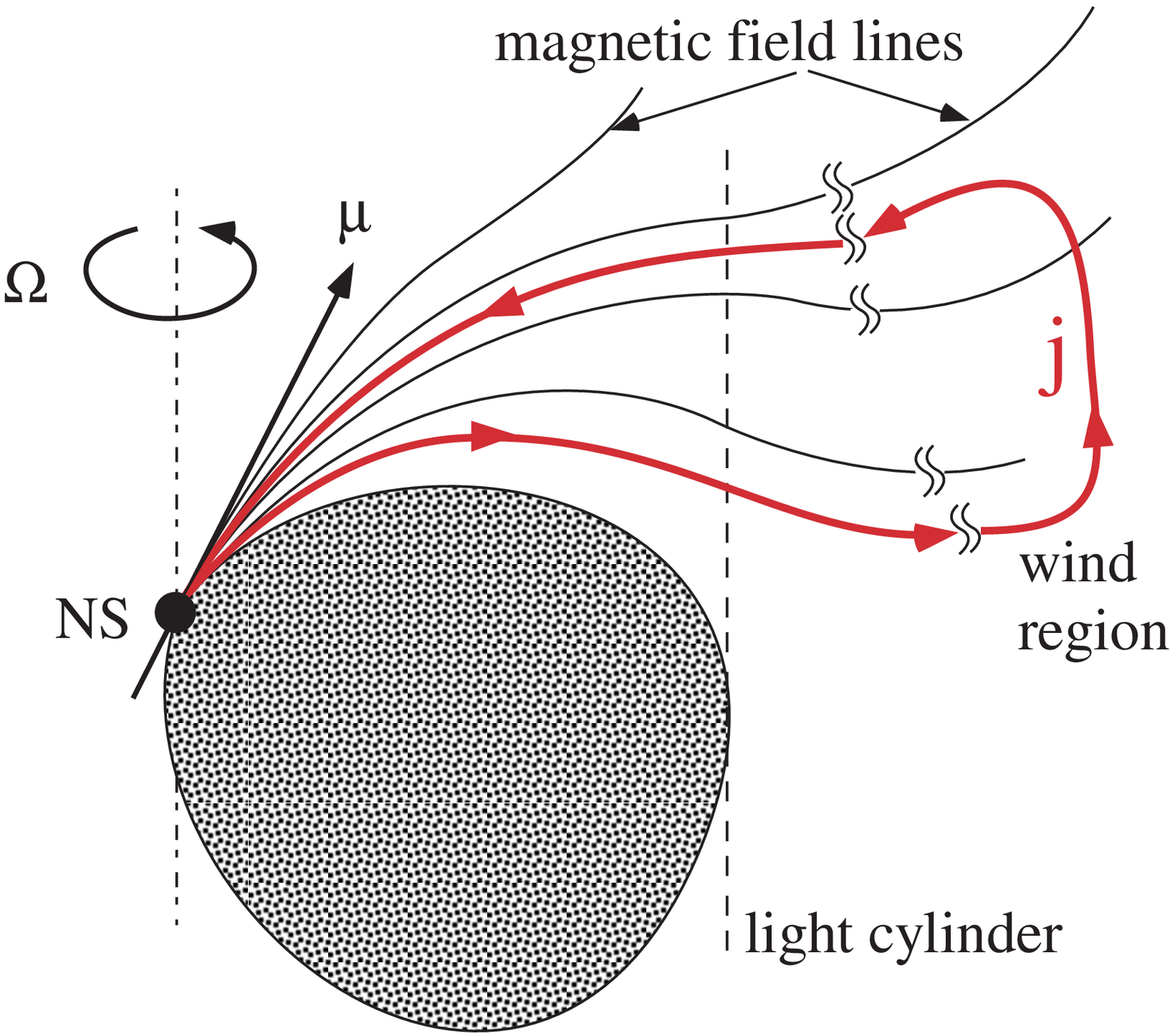}}
               \hspace*{0.5 truecm}
               {\epsfxsize=5.0truecm \epsfbox[-20 0 550 300]{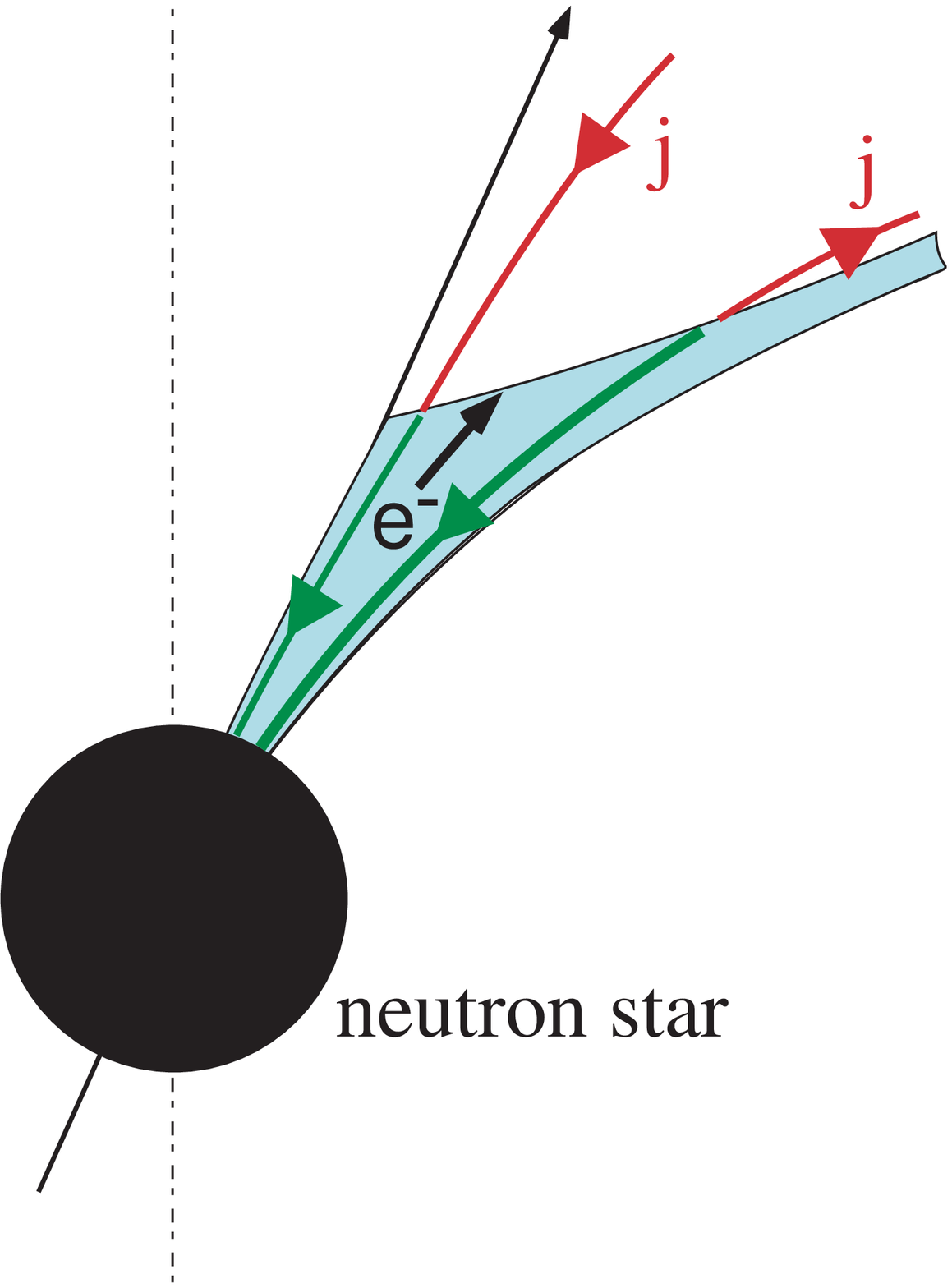}}}
\caption{
Sideview of pulsar magnetosphere.
{\it Left}:  two representative accelerator models in the open zone;
{\it center}: global electric current due to the EMF
              exerted on the spinning neutron star surface 
              when 
              $\mbox{\boldmath$\Omega$}\cdot\mbox{\boldmath$\mu$}>0$;
{\it right}: current derived in the inner-slot-gap model 
             (green arrow, see \S~6.4).
\label{fig:sidev}
}
\end{figure}

The pulsar magnetosphere can be divided into two zones:
The closed zone filled with a dense plasma corotating with the star,
and the open zone in which plasma flows along the open field lines
to escape through the light cylinder (left panel of fig.~\ref{fig:sidev}).
These two zones are separated by the the last-open magnetic field lines,
which become parallel to the rotation axis at the light cylinder.
Here, the light cylinder is defined as the surface where
the corotational speed of a plasma 
would coincide with the speed of light, $c$,
and hence its distance from the rotation axis is given by
$\rlc \equiv c/\Omega$.
If a plasma flows along the magnetic field line,
causality requires that the plasma should migrate outward 
outside of the light cylinder.
In all the pulsar emission models, 
particle acceleration takes place within the open zone.

For an aligned rotator (i.e., $\sin\inc=0$), 
open zone occupies the magnetic colatitudes 
(measured from the magnetic axis) that is less than $\sqrt{r_\ast/\rlc}$.
For an oblique rotator, even though the open-zone polar cap shape
is distorted, $\pi(r_\ast/\rlc)$ gives a good estimate 
of the polar cap area.
On the spinning neutron star surface, 
from the magnetic pole to the rim of the polar cap, 
an electro-motive force,
$\approx \Omega^2 B_\ast r_\ast^3/c^2 \approx 10^{16.5}$~V, is exerted.
This strong EMF causes the magnetospheric currents 
that flow outwards in the lower latitudes
and inwards along the magnetic axis (center panel in fig.~\ref{fig:sidev}).
The return current is formed at large-distances 
where Poynting flux is converted into kinetic energy of particles
or dissipated~\cite{Shibata97}.

Attempts to model the particle accelerator 
have traditionally concentrated on two scenarios:
Inner-gap models with emission altitudes of $\sim 10^4$cm
to several neutron star radii over a pulsar polar cap 
surface~\cite{Harding78,DH82,DH96,DS94,SDM95}
and outer-gap models with acceleration 
near the light cylinder~\cite{CHR86a,CHR86b,CR92,CR94,RY95}.
It is worth noting that average location of energy loss should take place
near the light cylinder so that the rotating neutron star may lose
enough angular momentum~\cite{Shibata95},
which indicates the existence of the inner gap must affect the 
electrodynamics in the outer magnetosphere.

It is widely accepted from phenomenological studies that
coherent {\it radio} pulsations are probably emitted from
the inner gap that is located near the magnetic 
pole.
Moreover, coherent radio pulsations provide the primary
channel for pulsar discovery 
(1560 pulsars as of April 2006, see ATNF pulsar 
 catalogue, http://www.atnf.csiro.au/research/pulsar/psrcat/).
However, it is, unfortunately, very difficult to reproduce them 
by considering plasma collective effects.
Therefore, in this review, 
I will focus on {\it incoherent high-energy}, non-thermal radiation,
which is the product of magnetospheric gaps.

It is worth mentioning other models than the inner and outer gap models.
A charge depletion can alternatively be a consequence of 
a stable charge void (i.e., no plasma region with 
 $\mbox{\boldmath$B$}\cdot\mbox{\boldmath$E$}\ne 0$)
in a neutron-star magnetosphere~\cite{Krause85a,Krause85b,Smith01}.
Even though their works provides important hints on electrostatics,
I do not intend to speculate on the dynamics of the formation of the void,
to take account of magnetospheric currents,
which are not considered in their works.
There is another model that radiation takes place in a pulsar wind 
at roughly 10 to 100 light cylinder radii from the star
as a result of the magnetic energy 
dissipation~\cite{Kirk01,Kirk02,Petri05},
which is a reexamination of the non-axisymmetric striped pulsar 
wind~\cite{Coron90,Michel94}.
However, to discuss the reconnection in a relativistic plasma
is beyond the scope of this brief review.

\section{Magnetospheric current determination}
\label{sec:current}
We briefly discuss how to determine the magnetospheric current.
Adopting the force-free limit (i.e., neglecting plasma inertia),
Ref.~\cite{CKF99} solved the Grad-Shafranov (GS) 
Eq.~\cite{Michel73,SW73,Okamoto74}
$(x^2-1)(\partial_x^2+\partial_z^2)P +(x+1/x)\partial_x P = j\,dj/dP$,
to obtain the magnetic flux function $P$,
which determines the fieldline configuration,
where $x$ and $z$
denote the distances from the rotation axis and the equatorial plane,
respectively, in $\rlc$ unit,
and $j$ the magnetospheric current 
on the poloidal (i.e., meridional) plane.
This work inaugurated numerous 
attempts~\cite{Goodwin04,Gruz05,Spit06,Timok06}
on solving the GS Eq. in recent several years,
and $k\approx 1+\sin^2\inc$ is suggested in Eq.~(\ref{eq:Ldip}),
by determining $j(P)$.
However, it is noteworthy that they determined $j$
by imposing a continuity of $P$ at the turning point $x=1$
(i.e., the light cylinder),
and by simply neglecting the highest-oder term at $x=1$
(including a regular Taylor expansion, which gives only one of the two
 independent solution bases if we suppress the $z$ dependence).
Instead, we can adopt a singular perturbation theory 
to stretch the region $x\approx 1$ by transforming $P=Q/(x-1)$.
Then there appears no singularity in the highest-order derivative 
term~\cite{Stix92}.
To find a particular solution,
we can, for example, introduce a regulator 
such that $P=Q/[x-1-\epsilon(z)]$, 
recovering the imaginary part by Fourier analysis
from the equations of force-free electrodynamics~\cite{Bland02,Spit06},
where $Q$ satisfies the GS Eq. without $z$ derivatives
and $\epsilon$ is a small complex function.
In this case, $x=1+\mbox{Re}(\epsilon)$ 
will give the Alfv\'enic separatrix surface~\cite{Bogov92,H98,H00},
after recovering particle inertia.
Any way, we must connect the interior solution 
with the asymptotic ones in $x<1$ and $x>1$.
It means that $j$, a free parameter in the force-free framework,
cannot be constrained by the continuity of $P$ at $x=1$,
but should be determined by the global requirement
including the dissipative region, which gives the electric load in the
current circuit 
within the sub-fast-magnetosonic region 
(extended to large distances in a magnetically dominated magnetosphere)
from causality requirement.

\section{Geometrical consideration}
\label{sec:geometry}
\begin{figure} 
  \centerline{ {\epsfxsize=4.2truecm \epsfbox[0 0 420 200]{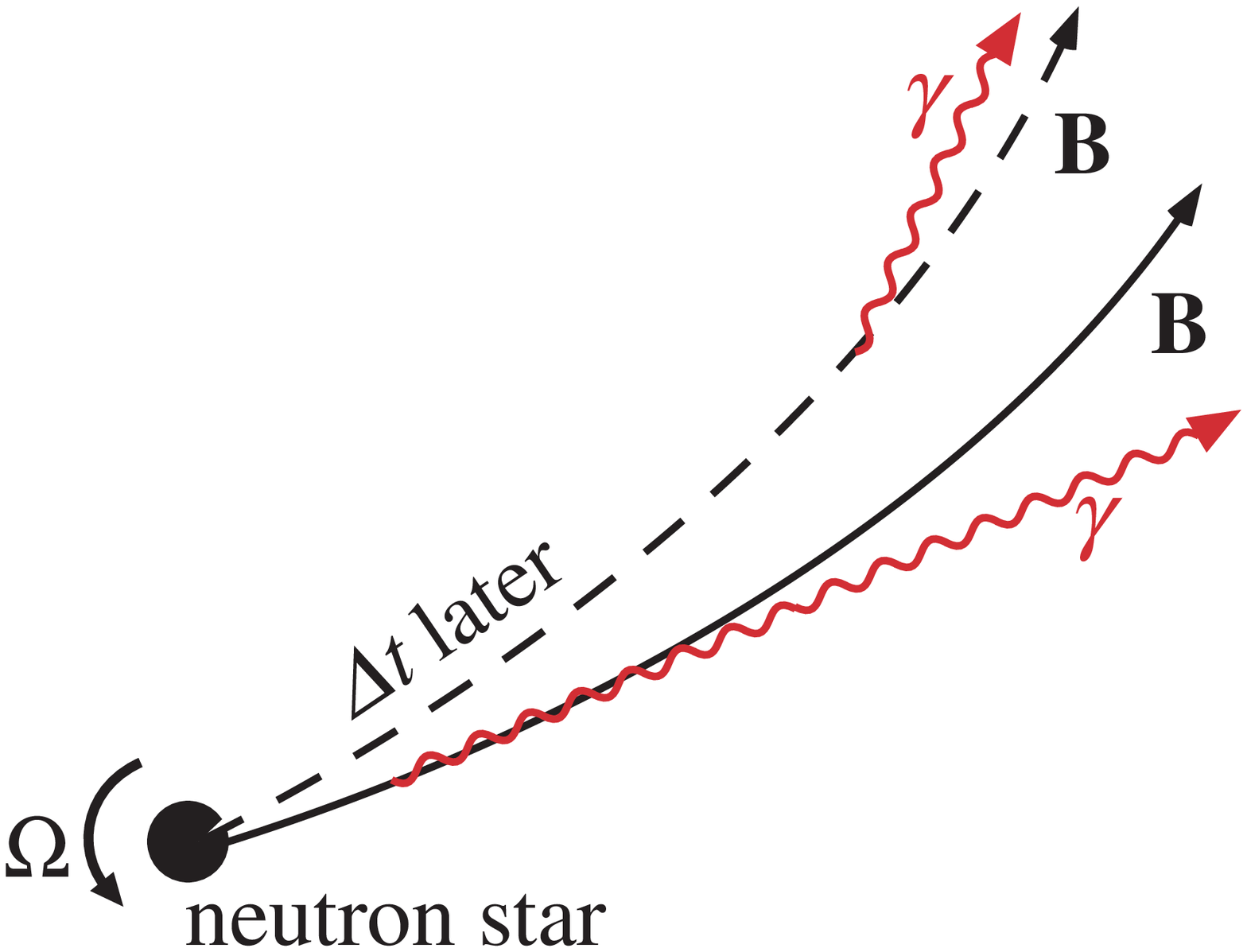}}
               \hspace*{1.0 truecm}
               {\epsfxsize=5.6truecm \epsfbox[0 0 500 300]{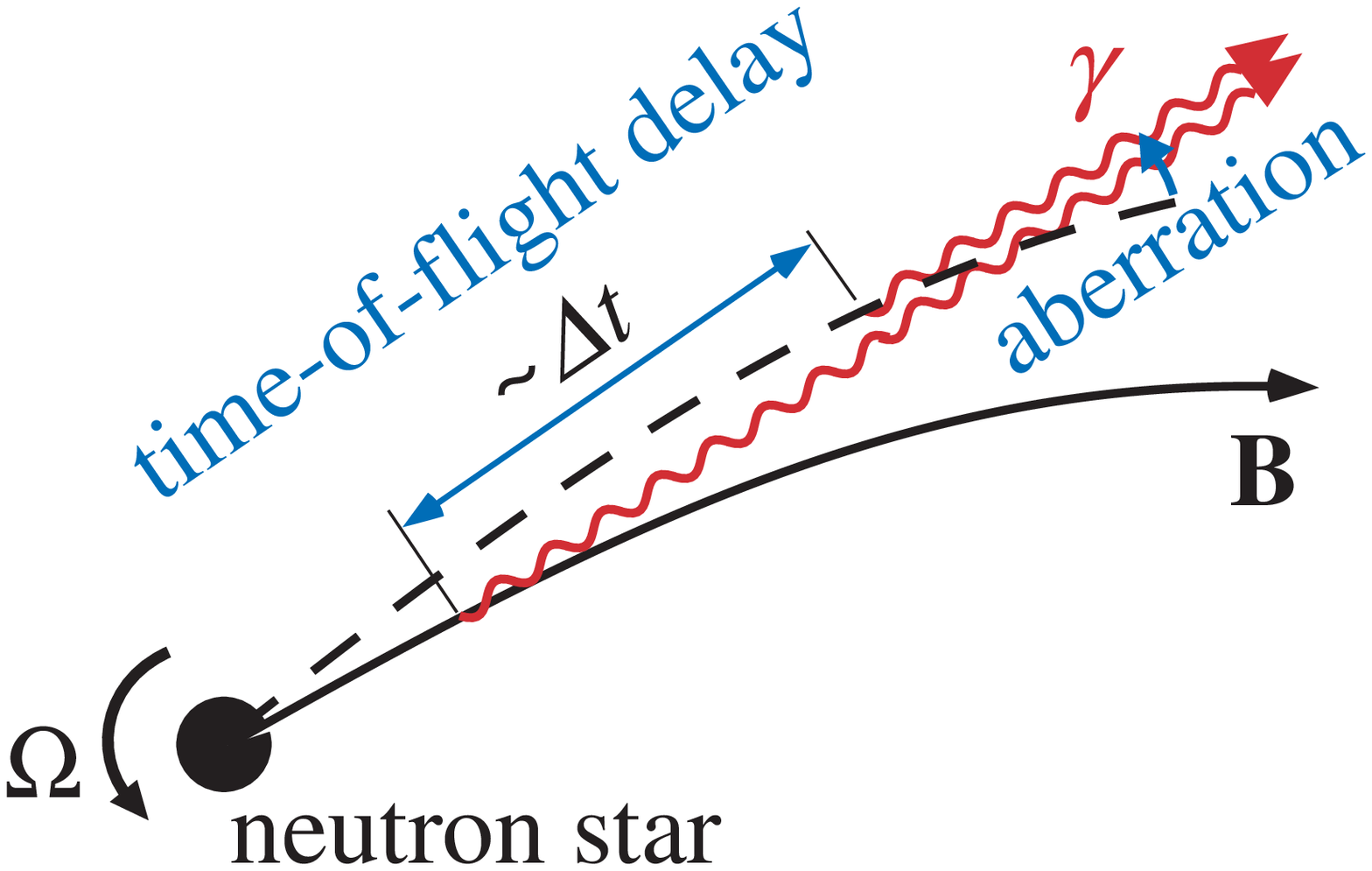}}}
\caption{
Photon propagation direction in a rotating neutron-star
magnetosphere (top view).
{\it Left}: Leading side emission, which forms weak photon intensity;
{\it right}: Trailing side emission, which forms a caustic
--- piling up of photons emitted at different altitudes.
\label{fig:caustics}
}
\end{figure}

In this section, we compare the emission geometry of
a few representative gap models.
For this purpose, we assume 
that the photons are outwardly emitted along the local 
magnetic field direction in the corotating frame of the magnetosphere.
Then the lightcurve morphologies are subject to the general relativistic
corrections (i.e., field distortion and light path bending near the star)
and the special relativistic corrections
(i.e., aberration of light, time-of-flight delays, and 
 field retardation near the light cylinder).
Aberration and time-of-flight delays leads to comparable phase shifts,
$\sim-\varpi/\rlc$ at distance $\varpi$ from the rotation axis.
On the leading side (left panel of fig.~\ref{fig:caustics}),
these phase shifts add up to spread photons emitted at various
altitudes over $0.4$ in phase.
On the other hand, on the trailing side (right panel), 
photons emitted later at higher altitudes catch up with those emitted
earlier at lower altitudes,
arriving at an observer within a small phase range $0.1$
and produce caustics~\cite{Morini83,RY95} 
in the phase plot and light curves~\cite{Cheng00,Dyks03}.

\begin{figure} [bp]
   \centerline{\hbox{\epsfxsize=16.0truecm
   \epsffile{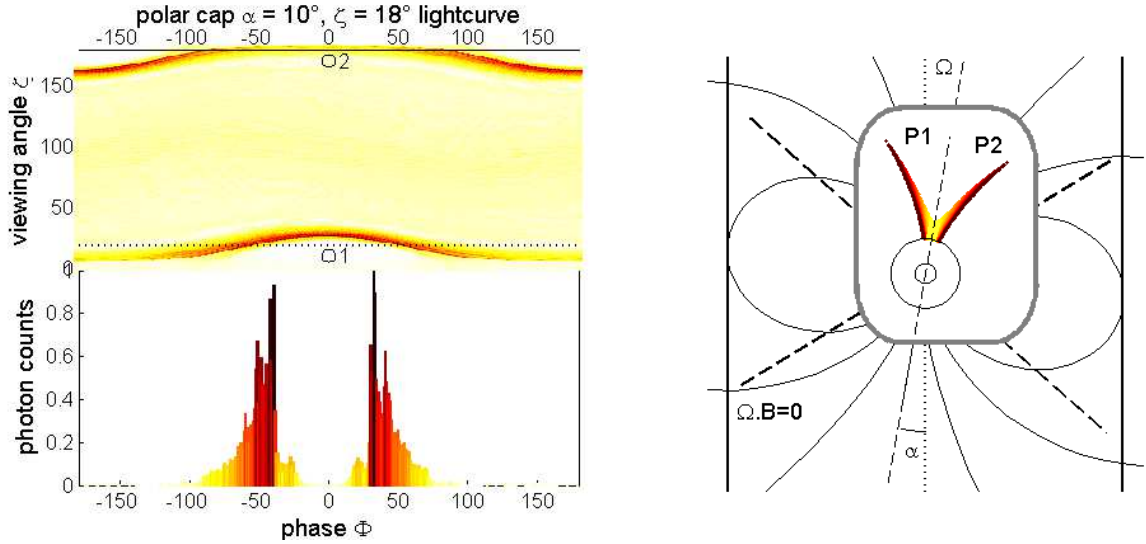}
     }
  }
\caption{
Phase plot, sample lightcurve, and a sketch of the accelerator location
for the inner-gap model,
for a typical inclination angle $\inc=10^\circ$.
the central zoom gives the gap extent relative to the star size.
the dashed lines outline the null surface.
the shading in the lightcurve and gap sketch is the same.
the phase plot illustrates the change in lightcurve as seen by
different observers and the aperture of the pulsed beams.
From Grenier and Harding (2006).
\label{fig:polar}
}
\end{figure}

A single inner-gap beam, which is emitted from the lower altitudes,
can produce a variety of pulse profiles.
The top left panel in Fig.~\ref{fig:polar} represents the phase plot,
--- photon intensity in the direction specified by 
$\Phi$ (pulse-phase) and 
$\zeta$ (observer's viewing angle with respect to the rotation axis).
An observer at $\zeta$ detects photon flux indicated by
the shade at each pulse phase (bottom left panel).
Because $\Ell$ fades away near the perfectly conducting edge of the
open zone, 
the gap is shorter near the pole and extends to higher altitude near the
rim (right panel).
Faint off-beam curvature radiation 
above the gap can be seen outside the main beam,
as the faintly shaded regions indicate\cite{HZ01}.
A great deal of effort has been made on the inner-gap 
model~\cite{Stur71,Ruder75,Miya97,Rudak99,Baring01};
however, one has to invoke a very small inclination angle
to reproduce the wide separation of the two peaks (fig.~\ref{fig:lcurvs}).
Thus, a high-altitude emission drew attention as an alternative 
possibility.

\begin{figure} 
   \centerline{\hbox{\epsfxsize=9.0truecm\epsffile{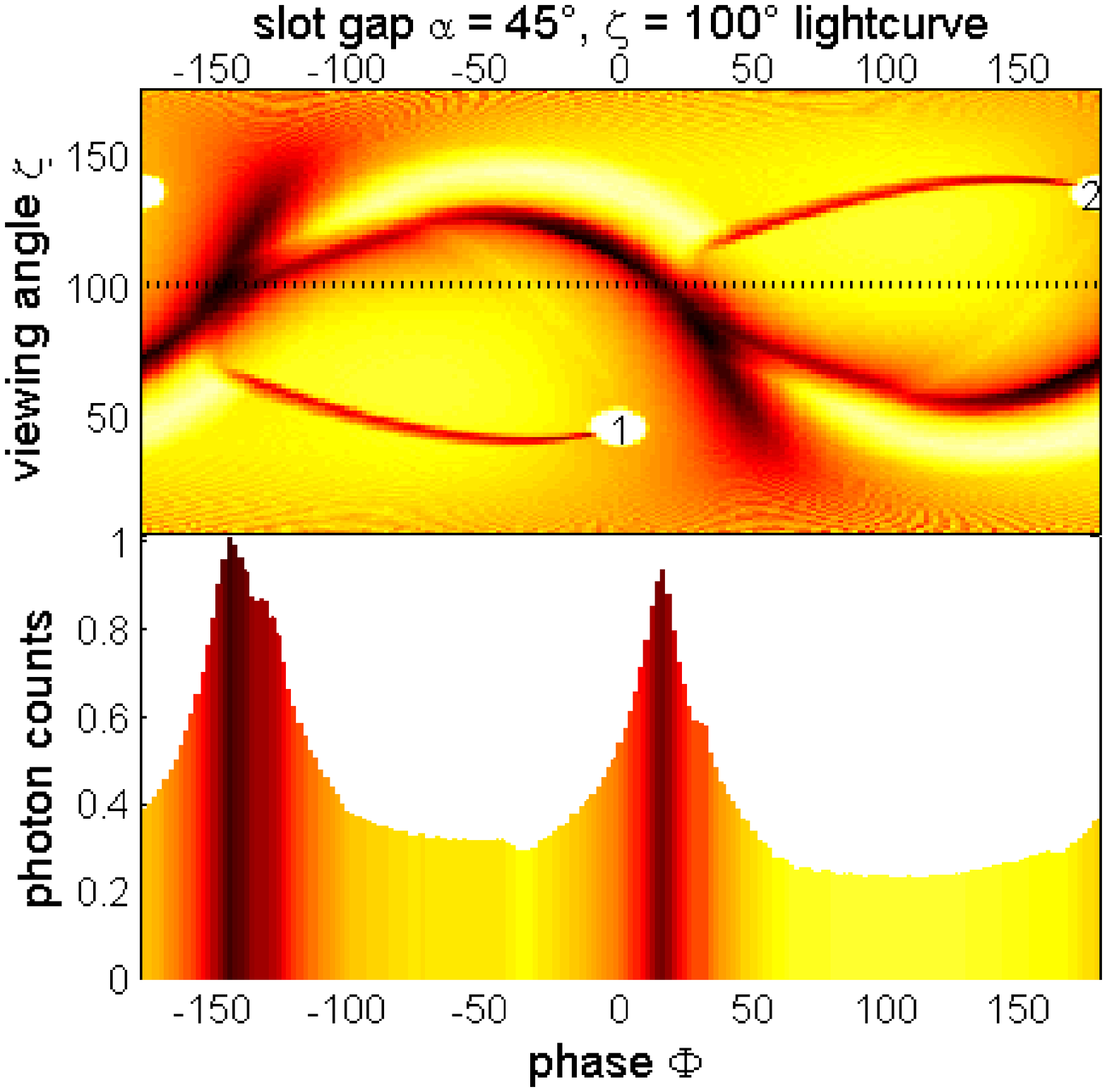}}
               \hbox{\epsfxsize=6.5truecm\epsffile{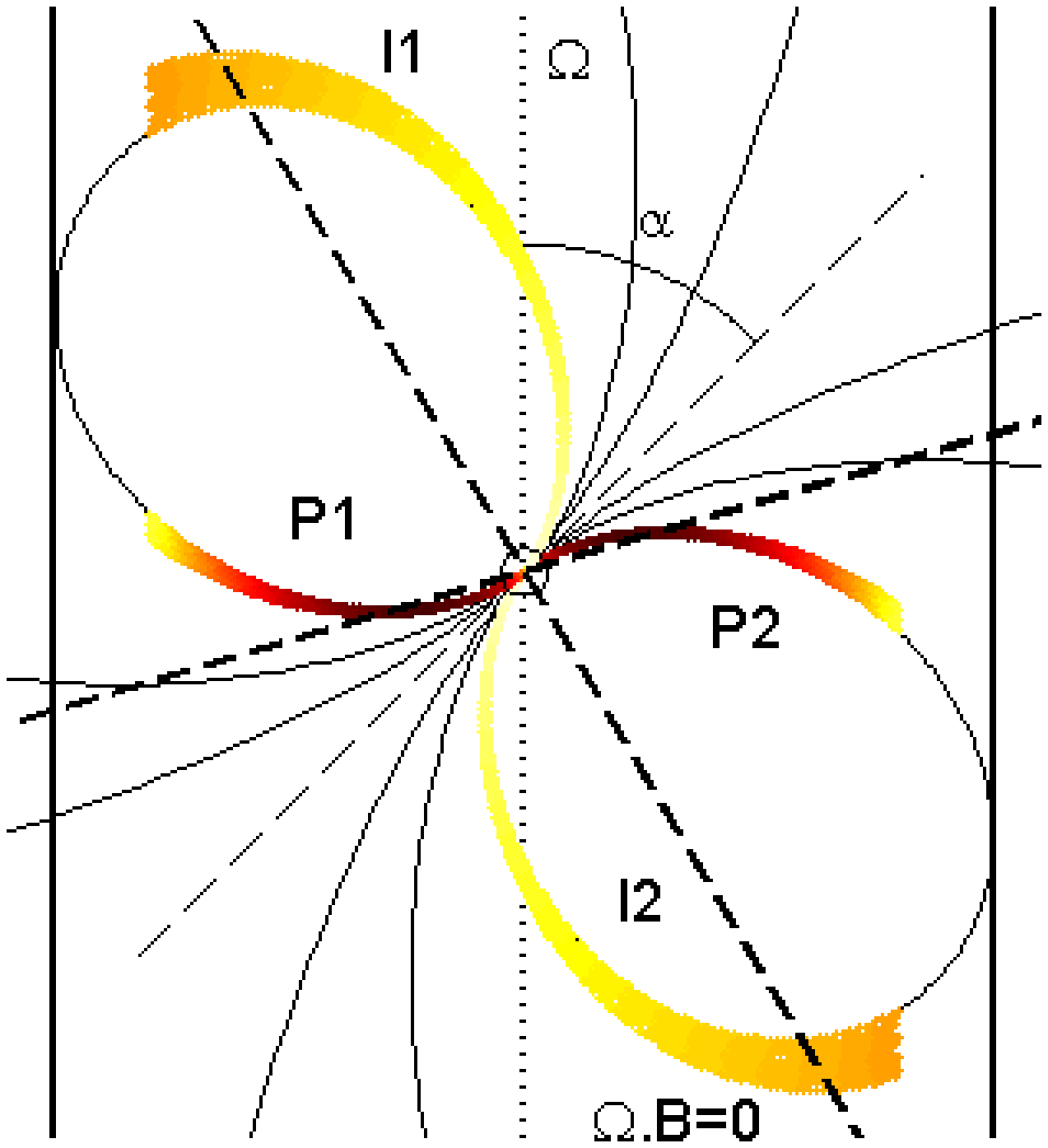}}
              }
\caption{
Same figure as fig.~1 for the slot-gap model,
for a typical inclination angle $\inc=45^\circ$.
From Grenier and Harding (2006).
\label{fig:PS}
}
\end{figure}

Assuming that the gap extends from the stellar surface 
to the light cylinder 
and that the photon emissivity is uniform within the gap,
Ref.~\cite{Dyks03} demonstrated the formation of the double peaks
arising from a crossing two caustics, 
each of which is associated with a different magnetic pole
(fig.~\ref{fig:PS}).
Subsequently, Ref.~\cite{Dyks04} examined polarization characteristics
and found that fast swings of the position angle and 
minima of polarization degree can be qualitatively reproduced 
within their two-pole caustic model.
This type of emission --- slot gap emission~\cite{GH06} ---
fills the whole sky and all phases in a lightcurve.
Most observers will catch emission from the two poles 
if $\inc \ge 30^\circ$
(for $45^\circ \le \zeta \le 125^\circ$ in fig.~\ref{fig:PS}).
The dark features show the accumulation of photons 
because of the trailing side caustics 
(e.g., the thick curve at $\zeta<130^\circ$ and $\vert\Phi\vert<50^\circ$),
and because of the overlap between the trailing side of pole 1
and the leading side of pole 2 near the light cylinder
(e.g., the thinner branch of the Y feature at 
 $50^\circ<\zeta<90^\circ$ and $50^\circ<\Phi<110^\circ$).
The main peaks come from the trailing side of each pole,
interpeak emission from the leading sides.

\begin{figure} [th]
   \centerline{\hbox{\epsfxsize=16.0truecm\epsffile{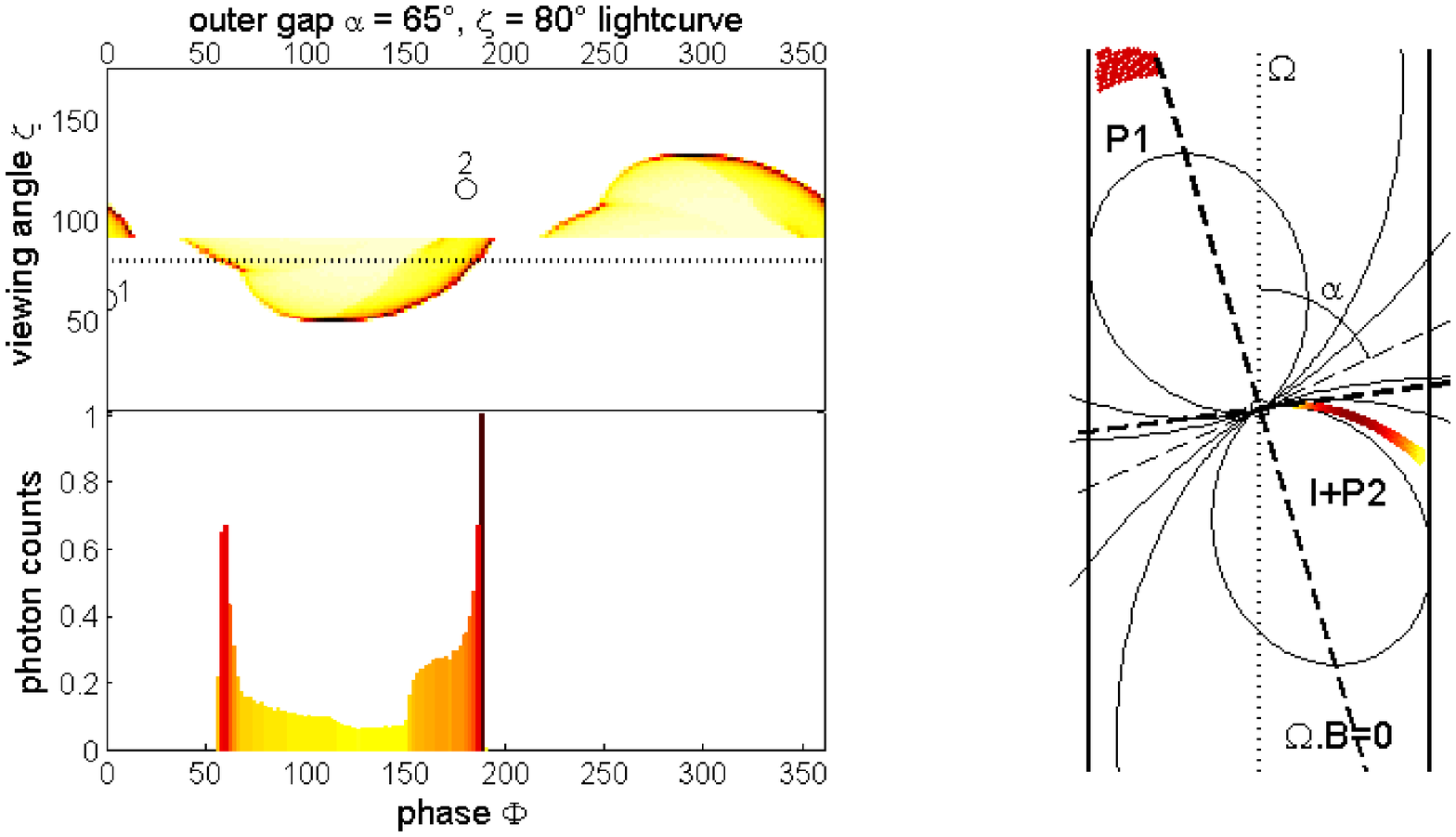}}
  }
\caption{
Same figure as fig.~1 for the outer-gap model,
for a typical inclination angle $\inc=65^\circ$.
From Grenier and Harding (2006).
\label{fig:outer}
}
\end{figure}

Geometrically speaking, an outer gap can be classified as
a subset of the slot gap extended above the null surface.
An observer can see, in this case, emission from the two outer gaps
associated with only one pole.
For example, in Fig.~\ref{fig:outer},
if $\zeta=80^\circ$, the dotted line shows that we observe the photons
emitted from the two outer gaps (shaded regions in the right panel)
associated with the pole 2.
There is no emission outside the sharp peak edges;
thus, the emission is invisible at $\zeta<30^\circ$ or $\zeta>150^\circ$
for any obliquity.
The first peak originate near $0.9\rlc$ 
(upper shaded region in the right panel), while the
second peak caustic is formed by the photons emitted from the trailing
field lines (lower shaded region in the right panel).

\section{Electrodynamical consideration}
\label{sec:electro}
Next, let us consider the physical basis of the geometrical models
discussed in \S~\ref{sec:geometry}.
To this aim, 
we have to derive the Poisson Eq. for the non-corotational potential,
which is applicable to arbitrary gap models.

\subsection{Poisson Equation for Electrostatic Potential}
\label{sec:poisson}
Around a rotating neutron star with mass $M$,
the background space-time geometry is given by~\cite{Lense18}
\begin{equation}
  ds^2= g_{tt} dt^2 + 2g_{t\varphi}dtd\varphi
       +g_{rr} dr^2 + g_{\theta\theta} d\theta^2
       +g_{\varphi\varphi} d\varphi^2,
  \label{eq:metric_1}
\end{equation}
where
\begin{equation}
  g_{tt}       
    \equiv \left( 1-\frac{r_{\rm g}}{r}\right) c^2, \,
  g_{t\varphi} 
    \equiv ac\frac{r_{\rm g}}{r}\sin^2\theta,
  \label{eq:metric_2}
\end{equation}
\begin{equation}
  g_{rr}   \equiv -\left( 1-\frac{r_{\rm g}}{r}\right)^{-1}, \,
  g_{\theta\theta}   \equiv -r^2, \,
  g_{\varphi\varphi} \equiv -r^2 \sin^2\theta;
  \label{eq:metric_3}
\end{equation}
$r_{\rm g}\equiv 2GM/c^2$ indicates the Schwarzschild radius,
and $a \equiv I\Omega/(Mc)$ parameterizes the stellar angular momentum.
At radial coordinate $r$, the inertial frame is dragged
at angular frequency 
$ \omega \equiv -g_{t\varphi}/g_{\varphi\varphi}
  = 0.15 \Omega I_{45} r_6{}^{-3} $,
where $I_{45} \equiv I/10^{45} \mbox{ erg cm}^2$, and 
$r_6 \equiv r_\ast/10\,\mbox{km}$.

Let us consider the Gauss's law,
\begin{equation}
  \nabla_\mu F^{t\mu}
  = \frac{1}{\sqrt{-g}}
    \partial_\mu \left[ \frac{\sqrt{-g}}{\rhowSQR}
                       g^{\mu\nu}(-g_{\varphi\varphi}F_{t\nu}
                               +g_{t\varphi}F_{\varphi\nu})
               \right]
  = \frac{4\pi}{c^2} \rho,
  \label{eq:Poisson_1}
\end{equation}
where $\nabla$ denotes the covariant derivative,
the Greek indices run over $t$, $r$, $\theta$, $\varphi$;
$\sqrt{-g}= \sqrt{g_{rr}g_{\theta\theta}\rhowSQR}=cr^2\sin\theta$ and
$\rhowSQR \equiv g_{t\varphi}^2-g_{tt}g_{\varphi\varphi}$,
$\rho$ the real charge density.
The electromagnetic fields observed by a distant static
observer are given by~\cite{Camen86a,Camen86b}
$ E_r=F_{rt}, \, E_\theta=F_{\theta t}, \, E_\varphi=F_{\varphi t}$,
$ B^r= (g_{tt}+g_{t\varphi}\Omega) F_{\theta\varphi}/\sqrt{-g}, \,
  B^\theta= (g_{tt}+g_{t\varphi}\Omega) F_{\varphi r}/\sqrt{-g}, \,
  B_\varphi= -\rhowSQR F_{r \theta}/\sqrt{-g}$,
where $F_{\mu\nu} \equiv \partial_\mu A_\nu-\partial_\nu A_\mu$ 
and $A_\mu$ denotes the vector potential.

Assuming that the electromagnetic fields are unchanged
in the corotating frame, we can introduce the 
non-corotational potential $\Psi$
such that
\begin{equation}
  F_{\mu t}+\Omega F_{\mu \varphi}
  = -\partial_\mu \Psi(r,\theta,\varphi-\Omega t),
  \label{eq:def_Psi}
\end{equation}
where $\mu= t,r,\theta,\varphi$.
If $F_{A t}+\Omega F_{A \varphi}=0$ ($A=r,\theta$) holds,
$\Omega$ is conserved along the field line.
On the neutron-star surface,
we impose $F_{\theta t}+\Omega F_{\theta\varphi}=0$ (perfect conductor) 
to find that the surface is equi-potential,
that is, 
$\partial_\theta \Psi= \partial_t \Psi +\Omega \partial_\varphi \Psi=0$
holds. 
However, in a particle acceleration region,
$F_{A t}+\Omega F_{A \varphi}$ deviates from $0$
and the magnetic field does not rigidly rotate.
The deviation is expressed in terms of $\Psi$, which gives
the strength of the acceleration electric field 
that is measured by a distant static observer as
\begin{equation}
  \Ell \equiv \frac{\mbox{\boldmath$B$}}{B}
              \cdot \mbox{\boldmath$E$}
       = \frac{B^i}{B}(F_{it}+\Omega F_{i\varphi})
       = \frac{\mbox{\boldmath$B$}}{B}
              \cdot (-\nabla\Psi),
  \label{eq:def_Ell}
\end{equation}
where the Latin index $i$ runs over spatial coordinates
$r$, $\theta$, $\varphi$.

Substituting Eq.~(\ref{eq:def_Psi}) into (\ref{eq:Poisson_1}),
we obtain the Poisson equation for the
non-corotational potential,
\begin{equation}
  -\frac{c^2}{\sqrt{-g}}
   \partial_\mu 
      \left( \frac{\sqrt{-g}}{\rhowSQR}
             g^{\mu\nu} g_{\varphi\varphi}
             \partial_\nu \Psi
      \right)
  = 4\pi(\rho-\rhoGJ),
  \label{eq:Poisson_2}
\end{equation}
where the general relativistic Goldreich-Julian charge density
is defined as
\begin{equation}
  \rhoGJ \equiv 
      \frac{c^2}{4\pi\sqrt{-g}}
      \partial_\mu \left[ \frac{\sqrt{-g}}{\rhowSQR}
                         g^{\mu\nu} g_{\varphi\varphi}
                         (\Omega-\omega) F_{\varphi\nu}
                 \right].
  \label{eq:def_GJ}
\end{equation}
If $\rho$ deviates from $\rhoGJ$ in any region,
$\Ell$ is exerted along $\mbox{\boldmath$B$}$.
In the limit $r \gg r_{\rm g}$, 
Eq.~(\ref{eq:def_GJ}) reduces to the
ordinary, special-relativistic expression~\cite{GJ69,Mestel71},
\begin{equation}
  \rhoGJ 
  \equiv -\frac{\mbox{\boldmath$\Omega$}\cdot\mbox{\boldmath$B$}}
               {2\pi c}
         +\frac{(\mbox{\boldmath$\Omega$}\times\mbox{\boldmath$r$})\cdot
                (\nabla\times\mbox{\boldmath$B$})}
               {4\pi c}.
  \label{eq:def_rhoGJ_1}
\end{equation}

Instead of ($r$,$\theta$,$\varphi$),
it is convenient to adopt the so-called \lq magnetic coordinates' 
($s$,$\theta_\ast$,$\varphi_\ast$)
such that $s$ denotes the distance along the magnetic field line,
$\theta_\ast$ and $\varphi_\ast$ the magnetic colatitude and 
the magnetic azimuth
of the point where the field line intersects the stellar surface.
Defining that $\theta_\ast=0$ corresponds to the magnetic axis
and that $\varphi_\ast=0$ to the plane on which both 
the rotation and the magnetic axes reside,
we obtain the following form of Poisson Eq.,
which can be applied to arbitrary magnetic field 
configurations~\cite{H06}:
\begin{eqnarray}
  &&
  -\frac{c^2 g_{\varphi\varphi}}{\rhowSQR}
  \left( g^{ss}\partial_s^2 
        +g^{\theta_\ast \theta_\ast} \partial_{\theta_\ast}^2
        +g^{\varphi_\ast \varphi_\ast}
            \partial_{\varphi_\ast}^2
        +2g^{s\theta_\ast} \partial_s \partial_{\theta_\ast}
        +2g^{\theta_\ast \varphi_\ast}
            \partial_{\theta_\ast} \partial_{\varphi_\ast}
        +2g^{\varphi_\ast s}
            \partial_{\varphi_\ast} \partial_s
  \right) \Psi
  \nonumber\\
  && -\left( A^s \partial_s
            +A^{\theta_\ast} \partial_{\theta_\ast}
            +A^{\varphi_\ast} \partial_{\varphi_\ast}
      \right) \Psi
  = 4\pi (\rho-\rhoGJ),
  \label{eq:BASIC_1}
\end{eqnarray}
\begin{equation}
  g^{i'j'} = g^{\mu\nu}\frac{\partial x^{i'}}{\partial x^\mu}
                       \frac{\partial x^{j'}}{\partial x^\nu}
           = g^{rr}\frac{\partial x^{i'}}{\partial r}
                   \frac{\partial x^{j'}}{\partial r}
            +g^{\theta\theta}
                   \frac{\partial x^{i'}}{\partial \theta}
                   \frac{\partial x^{j'}}{\partial \theta}
            -\frac{k_0}{\rhowSQR}
                   \frac{\partial x^{i'}}{\partial\varphi}
                   \frac{\partial x^{j'}}{\partial\varphi},
  \label{eq:def_mag1}
\end{equation}
\begin{equation}
  A^{i'} \equiv
  \frac{c^2}{\sqrt{-g}}
   \left\{ \partial_r \left[ \frac{g_{\varphi\varphi}}{\rhowSQR}
                             \sqrt{-g}g^{rr}
                             \frac{\partial x^{i'}}{\partial r}
                      \right]
          +\partial_\theta
                      \left[ \frac{g_{\varphi\varphi}}{\rhowSQR}
                             \sqrt{-g}g^{\theta\theta}
                             \frac{\partial x^{i'}}
                                  {\partial \theta}
                      \right]
    \right\}
   -\frac{c^2 g_{\varphi\varphi}}{\rhowSQR}
    \frac{k_0}{\rhowSQR}
    \frac{\partial^2 x^{i'}}{\partial\varphi^2},
  \label{eq:def_mag2}
\end{equation}
where $x^1=r$, $x^2=\theta$, $x^3=\varphi$,
$x^{1'}=s$, $x^{2'}=\theta_\ast$, and $x^{3'}=\varphi_\ast$.
The light surface, 
a generalization of the light cylinder,
is obtained by setting
$k_0 \equiv g_{tt}+2g_{t\varphi}\Omega+g_{\varphi\varphi}\Omega^2$
to be zero~\cite{Znaj77,Takah90}
Eq.~(\ref{eq:def_Ell}) gives 
$\Ell = -(\partial\Psi/\partial s)_{\theta_\ast,\varphi_\ast}$.
Magnetic field expansion effect~\cite{SAF78,MT92} is contained in 
$g^{\theta_\ast\theta_\ast}$,
$g^{\theta_\ast\varphi_\ast}$,
$g^{\varphi_\ast\varphi_\ast}$.

\subsection{Traditional Outer-gap Models}
\label{sec:outer}
Let us solve Eq.~(\ref{eq:BASIC_1})
for a vacuum case, $\rho=0$, on ($s$,$\theta_\ast$) plane,
neglecting $\varphi_\ast$ dependence.
Supposing that copious charged particles freely migrate along the
field lines outside of the gap,
it is natural to consider that the lower and upper boundaries
coincides with specific magnetic field lines.
Considering the field lines that thread the stellar surface on 
$\varphi_\ast=0$ plane,
we can express the lower and upper boundaries as 
$\theta_\ast=\theta_\ast^{\rm max}$ and 
$\theta_\ast=\theta_\ast^{\rm min}$, respectively.
A magnetic field line is specified by
the dimensionless quantity,
$h \equiv  (\theta_\ast^{\rm max}-\theta_\ast)/\theta_\ast^{\rm max}$;
$h=0$ and 
$h=h_{\rm m}\equiv 
 (\theta_\ast^{\rm max}-\theta_\ast^{\rm min})/\theta_\ast^{\rm max}$
corresponds to the lower and upper boundaries.
In the transversely thin limit (i.e., $h_{\rm m} \ll 1$),
we solve the elliptic-type 
partial differential Eq.~(\ref{eq:BASIC_1}) by imposing
$\Psi=0$ at $h=0$, $h=h_{\rm m}$ and $s=0$,
while $\partial_s \Psi=0$ at an artificially chosen
outer boundary ($s= 1.4\rlc$, say), which does not affect the solution
except for the vicinity of the light cylinder.
We assume $\mbox{\boldmath$\Omega$}\cdot\mbox{\boldmath$\mu$}>0$
throughout this paper.

As an example, we apply this method to the Crab pulsar,
which has been studied from radio, optical, X-ray to $\gamma$-ray wavelength
since its discovery~\cite{Stael68,Comel69}.
Its period and period derivative are
$P=33.0$~ms and $\dot{P}=4.20\times 10^{-13} \mbox{s s}^{-1}$.
If we adopt $k=2/3$ in Eq.~(\ref{eq:Ldip}), 
the observed $\dot{E}= -4.46\times 10^{38} \mbox{ergs s}^{-1}$ gives 
$\mu= 3.80\times 10^{30} (d/2\mbox{kpc})^2$.
From soft X-ray observations,
$180$~eV is obtained\cite{Tennant01} as the upper limit of the 
cooling neutron-star blackbody temperature, $kT$.

\begin{figure}[th]
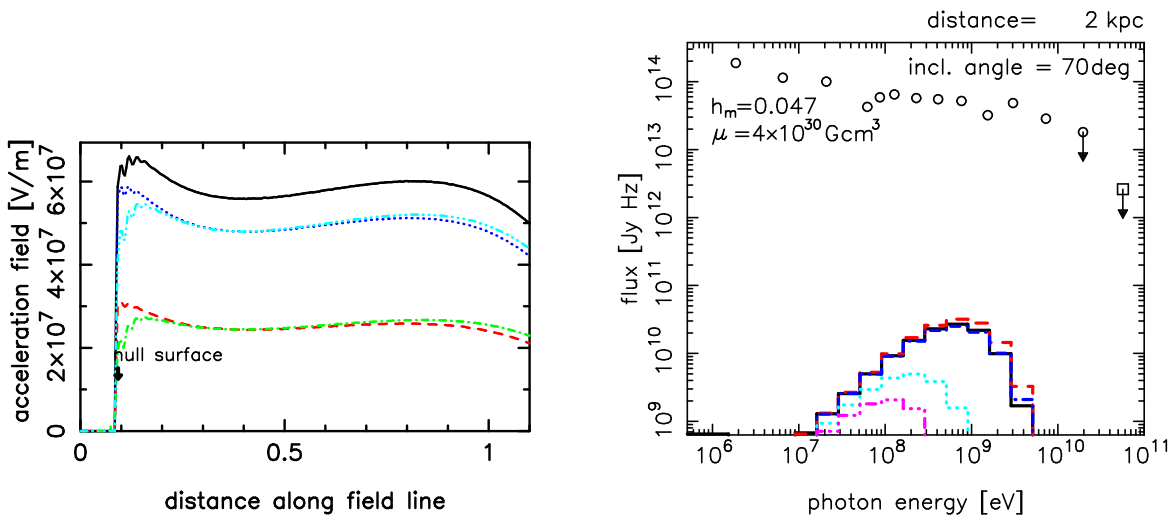

  \centerline{{\epsfxsize=8truecm \epsfbox[200 20 470 300]{f07a.eps}}
              {\epsfxsize=8truecm \epsfbox[200 20 470 300]{f07b.eps}}}
\vspace*{8pt}
\caption{
Traditional outer-gap solution obtained 
for the Crab pulsar with $\inc=70^\circ$ and $h_{\rm m}=0.047$.
{\it Left}: The field-aligned electric field
at discrete heights $h$
ranging from $2h_{\rm m}/16$, $5h_{\rm m}/16$, $8h_{\rm m}/16$,
$11h_{\rm m}/16$, $14h_{\rm m}/16$,
with dashed, dotted, solid, dash-dot-dot-dot, and dash-dotted curves,
respectively.
The abscissa indicates the distance along the field line
from the star in the unit of the light-cylinder radius.
The null surface position at the height $h=h_{\rm m}/2$ 
is indicated by the down arrow.
{\it Right}: Calculated phase-averaged spectra of the 
pulsed, outward-directed $\gamma$-rays.
The flux is averaged over the meridional emission angles between 
$44^\circ$ and $58^\circ$ (solid),
$58^\circ$ and $72^\circ$ (dashed),
$72^\circ$ and $86^\circ$ (dash-dotted), 
$86^\circ$ and $100^\circ$ (dotted), 
$100^\circ$ and $114^\circ$ (dash-dot-dot-dotted),
from the magnetic axis
on the plane in which both the rotational and magnetic axes reside.
\label{fig:E70sub}
}
\end{figure}

Adopting $kT=100$~eV, $\mu=4.0 \times 10^{30} \mbox{G cm}^3$, and 
the magnetic inclination $\inc=70^\circ$,
which is more or less close to the value ($65^\circ$) suggested 
by a three-dimensional analysis in the traditional outer gap 
model (CHR)~\cite{Cheng00},
we obtain a nearly vacuum solution (fig.~\ref{fig:E70sub})
for a geometrically thin case $h_{\rm m}=0.047$.
In the left panel, we present $\Ell(s,h)$ at five discrete heights.
The solutions become similar to the vacuum one
obtained in CHR.
For one thing, 
the inner boundary is located slightly inside of the null surface.
What is more, $\Ell$ maximizes at the central height, $h=h_{\rm m}/2$,
and remains roughly constant in the entire region of the gap.
The solved $\Ell$ distributes almost symmetrically with respect to the
central height; for example, the dashed and dash-dotted curves
nearly overlap each other.
The gap has no outer termination within the light cylinder.
Since the inner boundary coincides with the place where
$\Psi$ vanishes, the region between the star and the inner boundary
has $\Psi>0$.
Therefore, negative charges pulled from the stellar surface
with $\Ell\equiv -\partial_s\Psi<0$ populate 
only inside of the inner boundary, in which $\Psi<0$ holds.
Similar solutions are obtained for a thiner gap,
$h_{\rm m}<0.047$,
even though there appears a small $\Ell$ peak
near the null surface, which is less important.

Refs.~\cite{CHR86a,CHR86b} 
first considered this kind of vacuum solutions
and suggested two outer gaps can be formed for each magnetic pole
(i.e., the upper and lower shaded regions in the right panel
 of fig.~\ref{fig:outer}).
Considering a fan beam 
(instead of a pencil beam in the inner-gap model
 or a funnel beam in the inner-slot-gap model) as the emission morphology,
they discussed the formation of cusped photon peak
as shown in Fig.~\ref{fig:outer}.
Extending this morphological emission model,
Stanford group~\cite{RY95} discussed the observed properties of 
individual $\gamma$-ray pulsars, 
their radio to $\gamma$-ray pulse offsets,
and the radio- vs. $\gamma$-ray detection probabilities.
Assuming $\Ell \propto s^{-1}$ and
a power-law energy distribution accelerated $e^\pm$'s,
Ref.~\cite{Romani96} estimated the evolution of
high-energy flux efficiencies and beaming fractions
to discuss the detection statistics.
Another group in Hong Kong~\cite{ZC97} examined the minimum trans-field
thickness, $h_{\rm m}$, of the gap,
imposing that the $\gamma$-$\gamma$ pair creation criterion is met.
They estimated the soft photon field emitted from the heated
polar-cap surface by the bombardment of gap-accelerated charged particles,
adopting essentially the same $\Ell$ solution as Fig.~\ref{fig:E70sub}.
Extending this work,
Ref.~\cite{Cheng00}
developed a three-dimensional outer magnetospheric gap model
to examine the double-peak light curves with strong inter-pulse emission
(fig.~\ref{fig:outer}),
and estimated phase-resolved $\gamma$-ray spectra 
by assuming that the charged particles are accelerated 
to the Lorentz factors
at which the curvature radiation-reaction force balances with 
the electrostatic acceleration.

The outer gap models of these two groups 
have been successful in explaining the observed light curves,
particularly in reproducing the wide separation of the two peaks
(fig.~\ref{fig:lcurvs}),
without invoking a very small inclination angle
(as in inner-gap models).
However, if we solve Eq.~(\ref{eq:BASIC_1})
self-consistently with the particle and $\gamma$-ray Boltzmann 
Eqs.\cite{H06},
we find that the $\gamma$-ray flux obtained for
$h_{\rm m}<0.047$ (i.e., traditional outer-gap models)
is insufficient (right panel of fig.~\ref{fig:E70sub}).
Thus, we have to consider a transversely thicker gap,
which exerts a larger $\Ell$ because of the less-efficient
screening due to the two zero-potential walls
at $h=0$ and $h_{\rm m}$.
If the created current increases due to the increased $h_{\rm m}$,
the gap inner boundary deviates the null surface and shifts inwards
to touch the stellar surface at last\cite{HHS03}.
Therefore, 
we will analytically prove this behavior in \S~\ref{sec:innerBD}
then move on to the main task.

\subsection{Inner boundary position}
\label{sec:innerBD}
In the transversely thin (i.e., vacuum) limit,
$\theta_\ast$ derivatives dominate in Eq.~(\ref{eq:BASIC_1}) to give
$-(B/B_\ast)r_\ast^{-2}\partial^2\Psi/\partial\theta_\ast^2
 \approx -4\pi\rhoGJ$.
Noting that $\rhoGJ>0$ and $\partial_s\rhoGJ>0$ hold within the gap, 
and that $\Psi=0$ at both $\theta_\ast=\theta_\ast^{\rm min}$
and $\theta_\ast=\theta_\ast^{\rm max}$,
we find $\Psi<0$ and $\Ell\equiv -\partial_s\Psi>0$.
That is, the negativity of $\rho-\rhoGJ$ results in 
a positive $\Ell$ in the gap, if $\theta_\ast$ derivatives dominate.
Then, what happens near the inner boundary where
$s$ derivatives become important.
To avoid the reversal of the sign of $\Ell$,
we find that 
$\partial\Ell/\partial s=-\partial^2\Psi/\partial s^2
 \approx -4\pi\rhoGJ>0$ must be satisfied.
Therefore, we can conclude that the inner boundary should be
located slightly inside of the null surface where $\rhoGJ$ changes sign.

The same argument holds for a non-vacuum gap. 
At the inner boundary, $\partial_s\Ell\approx 4\pi(\rho-\rhoGJ)>0$
leads to
\begin{equation}
  -\frac{\rhoGJ}{B}
  \sim \frac{\Omega}{2\pi c}\left( 1-\frac{\omega}{\Omega} \right)
       \frac{B_z}{B}
  > \frac{\vert\rho\vert}{B}
  \sim \frac{\Omega}{2\pi c},
  \label{eq:inBDcond}
\end{equation}
so that $\Ell$ may have a single sign.
It follows that the polar cap, where $B_z \sim B_z^\ast$ holds,
is the only place for the inner boundary of the \lq outer' gap 
to be located,
if the created particle number density is comparable to the
Goldreich-Julian value at the surface,
namely, $\rho \sim \Omega B_z^\ast/(2\pi c)$.
Such a non-vacuum gap must extend from the polar cap surface
(not from the null surface as traditionally assumed)
to the outer magnetosphere.
On these grounds, 
we have to merge the inner- and outer-gap models,
which have been separately considered so far.

\subsection{Inner-slot-gap models}
\label{sec:slog}
In traditional inner-gap models,
the predicted beam size of radiation emitted 
near the stellar surface
is too small to produce the wide pulse profiles (fig.~\ref{fig:lcurvs}).
Seeking the possibility of a wide hollow cone of high-energy radiation
due to the flaring of field lines,
Ref.~\cite{A83} first examined the particle acceleration
at the high altitudes along the last open field line.
This type of accelerator, or the slot gap, 
forms because the pair formation front, which screens $\Ell$, 
occurs at increasingly higher altitude as the magnetic colatitude
approaches the edge of the open field region~\cite{AS79}.
MH03~\cite{MH03} extended this argument
by including two new features:
acceleration due to space-time dragging, and
the additional decrease of $\Ell$ at the edge of the gap
due to the narrowness of the slot gap.
Moreover, to give the physical basis of 
(purely geometrical) two-pole caustic model (\S~\ref{sec:geometry}),
MH04a,b~\cite{MH04a,MH04b}
matched the high-altitude slot gap solution 
to the solution obtained at lower altitudes by MH03,
and found that the residual $\Ell$ is 
small and constant,
but still large enough at all altitudes to maintain 
the energies of electrons, 
which are extracted from the star, above 5~TeV.

We should notice here that their inner-slot gap model
is an extension of the inner-gap model
into the outer magnetosphere,
assuming a space-charge-limited flow (SCLF)
that the plasma flowing in the gap consists of only 
the charges extracted from the stellar surface. 
This assumption is self-consistently satisfied in their model,
because pair creation in the extended slot gap occurs at a
reduced rate.
However, the SCLF leads to the electric current 
that is opposite to the global current flow patterns,
if the gap exists near the last-open field line
(right panel of fig.~\ref{fig:sidev}).

\subsection{Modern outer-gap models
   --- super Goldreich-Julian current with ion emission from stellar surface}
\label{sec:outer2}
We are, therefore, motivated by the need to contrive 
an accelerator model that predicts a consistent current direction
with the global requirement. 
To this aim, it is straightforward to extend recent outer-gap models,
which predict opposite $\Ell$ to polar-cap models,
into the inner magnetosphere.
Extending the one-dimensional 
studies~\cite{Beskin92,HO98,HS99a,HS99b,HS99c,HHS03},
Refs.~\cite{Takata04,Takata06}
solved (a simpler form of) Eq.~(\ref{eq:BASIC_1}) 
to reveal that the gap inner boundary is located inside of the
null surface owing to the pair creation within the gap,
assuming that the particle motion immediately saturates
in the balance between electric and radiation-reaction forces.

\begin{figure}[th]
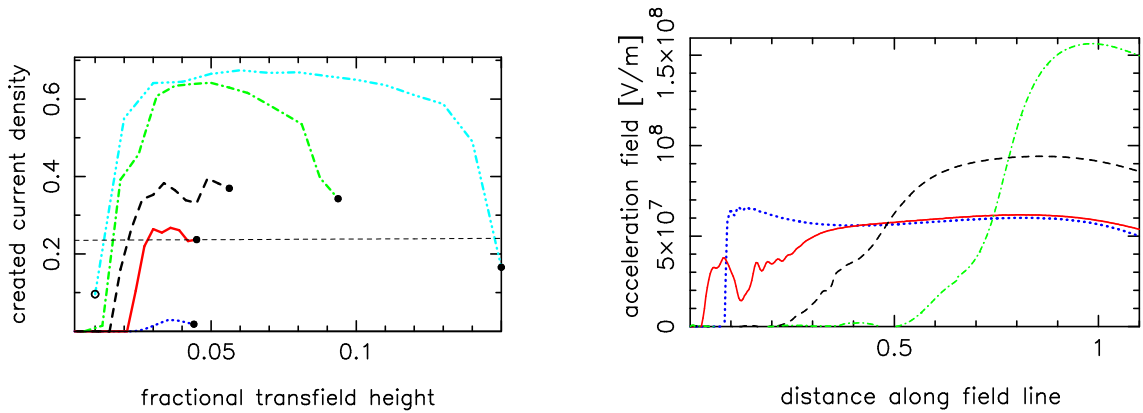

  \centerline{{\epsfxsize=8truecm \epsfbox[200 20 470 300]{f08a.eps}}
              {\epsfxsize=8truecm \epsfbox[200 20 470 300]{f08b.eps}}}
\caption{
Modern outer-gap solution obtained 
for the Crab pulsar with $\inc=70^\circ$ and $h_{\rm m}\ge0.048$:
The dotted, solid, dashed, dash-dotted, and dash-dot-dot-dotted curves
corresponds to 
$h_{\rm m}=0.047$, $0.048$, $0.060$, $0.100$, and $0.160$, respectively.
{\it Left}: Created current density $j_{\rm e}$ (in unit of $\Omega B/2\pi$)
as a function of the transfield thickness $h$.
If $j_{\rm e}$ appears below (or above) the dashed line,
$\vert c\rhoGJ/(\Omega B/2\pi) \vert_{s=0}$,
the solution is sub- (or super-) GJ current.
{\it Right}: $\Ell(s,h_{\rm m}/2)$.
\label{fig:C70d}
}
\end{figure}

\begin{figure}
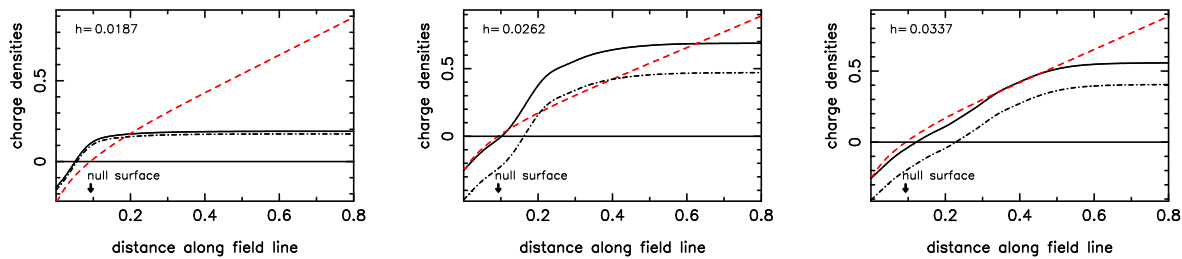

  \centerline{{\epsfxsize=5.3truecm \epsfbox[200 20 470 300]{f09a.eps}}
              {\epsfxsize=5.3truecm \epsfbox[200 20 470 300]{f09b.eps}}
              {\epsfxsize=5.3truecm \epsfbox[200 20 470 300]{f09c.eps}}}
\caption{
Modern outer-gap solution of the 
total (solid), created (dash-dotted), and Goldreich-Julian (dashed)
charge densities in $\Omega B(s,h) /(2\pi c)$ unit,
for $\inc=70^\circ$ and $h_{\rm m}=0.060$ 
at three transfield heights, $h$.
Because of an ion emission from the stellar surface,
the total charge density deviates from the created one.
\label{fig:C70three}
}
\end{figure}

Extending Ref.~\cite{Takata04,Takata06}
by solving the Lorentz-factor and pitch-angle evolution of individual
$e^\pm$'s,
H06~\cite{H06} solved the gap electrodynamics for the Crab pulsar.
For $h_{\rm m}>0.047$,
the created current density $j_{\rm e}$ becomes
super Goldreich-Julian in the sense that
$\rho<\rhoGJ<0$ holds at the inner boundary 
(left panel of fig.~\ref{fig:C70d}).
The predicted $\gamma$-ray flux is much larger than the sub-GJ case
of $h_{\rm m}\le 0.047$ (i.e., traditional outer-gap solution).
Moreover, the copious pair creation leads to a substantial screening
of $\Ell$ in the inner region, as the right panel shows.
In this screening region, $\rho(s,h)$ distributes
(fig.~\ref{fig:C70three}) so that $\Ell$ may virtually vanish.
Because of the ion emission from the stellar surface,
the total charge density $\rho$ is given by
$\rho=\rho_{\rm e}+\rho_{\rm ion}$,
where $\rho_{\rm e}$ denotes the sum of positronic and electronic
charge densities,
while $\rho_{\rm ion}$ does the ionic one.
We should notice here that even if $\Ell\approx 0$ occurs
by the discharge of created pairs in most portions of the gap,
the negative $\rho_{\rm e}-\rhoGJ$ inevitably exerts a strong 
positive $\Ell$ at the surface (see H06 for details),
thereby extracting ions from the surface
for the solution with super-GJ current.
On these grounds, we can regard this modern outer-gap solution
as a mixture of the traditional inner-gap model,
which extracts electrons from the surface with $\Ell<0$,
and the traditional outer-gap model,
which exerts positive $\Ell$ because of the negativity of $\rho-\rhoGJ$.

Most of the inward-directed $\gamma$-rays 
materialize outside of the gap, leading to a pair cascade
by colliding with the surface X-rays.
Theses pairs quickly lose initial inward momenta 
by a strong synchrotron radiation to become non-relativistic,
then are resonantly scattered by the surface X-rays,
and finally form a pulsar wind.
It follows that $8\times10^{38}$ particles are created per second,
which appears less than the constraints
that arise from consideration of
magnetic dissipation in the wind zone\cite{Kirk03}
($10^{40}\,\mbox{s}^{-1}$),
and of Crab Nebula's radio synchrotron emission\cite{Arons04}
($10^{40.5}\,\mbox{s}^{-1}$).

I summarize accelerator models/theories in Table~\ref{tbl:1}.
The second and third columns represent the 
spatial dimension of the model and the sign of exerted $\Ell$.

 \begin{table}[ht]
 \begin{center}
 \begin{small}
 \caption{Electrodynamical models
     of high-energy emission from pulsar magnetospheres
     \label{tbl:1}}
{\begin{tabular}{clllcllll}
\hline\hline
ref.  & D               & $\Ell$               & $\rho$ in 
      & acceleration    
      & Lorentz factor  & pitch angle          & $\gamma$-ray    
      & pulse         \\
      &                 &                      & Poisson eq.
      & electric field
      & dependence 
                        & dependence 
                                               & spectrum        
      & profiles      \\
\hline
\cite{MH04a}
      &                 &                      & assumed in
      & solved for
      & mono-energetic  & not                  & not
      &               \\
\cite{MH04b}
      & 3               & $-$                  & $r>5r_\ast$
      & assumed $\rho$
      & $\Gamma^4/R_{\rm c}\propto \Ell$ 
                        & solved               & solved  
      & examined      \\
\hline
\cite{RY95}
      &                 &                      & 
      & assumed  
      & assumed to be   & not                  & 
      &               \\
\cite{Romani96}
      & 3               & $+$                  & vacuum
      & $\Ell\propto r^{-1}$
      & power-law
                        & solved           & solved
      & examined      \\
\hline
      &                 &                      & 
      & solved to be
      & mono-energetic  & not                  & 
      &               \\
\cite{Cheng00}
      & 3               & $+$                  & vacuum
      & $\Ell\propto r^{0}$
      & $\Gamma^4/R_{\rm c}\propto \Ell$ 
                        & solved               & solved
      & examined      \\
\hline
      &                 &                      & solved from
      & solved from
      & solved from     & not                  & 
      & not           \\
\cite{HHS03}
      & 1               & $+$                  & pair prod.
      & Poisson eq.
      & Boltzmann eq.   & solved               & solved
      & examined      \\
\hline
      &                 &                      & solved from
      & solved from
      & mono-energetic  & not                  & 
      & not           \\
\cite{Takata04}
      & 2               & $+$                  & pair prod. 
      & Poisson eq.
      & $\Gamma^4/R_{\rm c}\propto \Ell$ 
                        & solved               & solved
      & examined      \\
\hline
      &                 &                      & solved from
      & solved from
      & solved from     & solved from          & 
      & not           \\
\cite{H06}
      & 2               & $+$                  & pair prod.
      & Poisson eq.
      & Boltzmann eq.   & Boltz. eq.           & solved
      & examined      \\
\hline\hline
\end{tabular}}
 \end{small}
 \end{center}
\end{table}

\subsection{Inner-slot-gap model vs. modern outer-gap model}
\label{sec:cf}
It is worth comparing the results of H06 
with the inner-slot-gap model proposed by MH03, MH04a,b,
who obtained a quite different solution 
(e.g., negative $\Ell$ in the gap).
The only difference is, essentially, the transfield thickness of the gap.
Estimating the transfield thickness to be
$\Delta l_{\rm SG} \sim h_{\rm m} r_\ast \sqrt{r/\rlc}$,
which is a few hundred times thinner than H06,
they extended the solution near the polar cap surface~\cite{MT92} 
into the higher altitudes (towards the light cylinder).
Because of this very small $\Delta l_{\rm SG}$,
emitted $\gamma$-rays do not efficiently materialize within the gap;
as a result, the created and returned positrons from the higher altitudes
do not break down the original assumption of SCLF near the stellar surface.

To avoid the reversal of $\Ell$ in the gap
(from negative near the star to positive in the outer magnetosphere),
or equivalently,
to avoid the reversal of the sign of the effective charge density,
$\rho_{\rm eff} = \rho-\rhoGJ$, along the field line,
MH04a,b assumed that
$\rho_{\rm eff}/B$ nearly vanishes and
remains constant above a few neutron star radii.
Because of this assumption, $\Ell$ is suppress at a very small value,
which reduces pair creation significantly.
To justify this $\rho/B$ distribution,
they proposed that $\rho$ should grow by 
the {\it cross field motion} of charges due to toroidal 
forces~\cite{Mestel85},
and that $\rho_{\rm eff}/B$ is a small constant so that 
$c\rho_{\rm eff}/B$ may not exceed the flux of the 
emitted charges from the star,
which ensures the {\it equipotentiality} of the slot-gap boundaries.
Because of the very thin gap, $\Delta l_{\rm SG} \sim r_\ast/20$,
the cross-field motion becomes, indeed, important.
However, the assumption that $\rho_{\rm eff}/B$ is a small positive
constant may be too strong, because it is only a sufficient condition
of equipotentiality.

On the other hand, H06 considered a thicker gap,
$\Delta l_{\rm SG} \sim 0.5 h_{\rm m} \rlc$,
in which the cross-field motion is negligible.
In the inner magnetosphere, 
$\rho_{\rm eff}/B$ becomes approximately a negative constant,
owing to the discharge of the copiously created pairs;
as a result, a positive $\Ell$ is exerted.
For a super-GJ solution, we obtain
$j_{\rm e}+j_{\rm ion} \sim 0.9 > \rho_{\rm eff}/(\Omega B/2\pi)$,
which guarantees the equipotentiality.

In short, whether the gap solution becomes 
MH04a way 
(with a negative $\Ell$ as an outward extension of the inner-gap model)
or H06 way 
(with a positive $\Ell$ as an inward extension of the outer-gap model)
entirely depends on the transfield thickness and on 
the resultant $\rho_{\rm eff}/B$ variation along the field lines.
If $\Delta l_{\rm SG} \sim r_\ast/10$ holds in the outer magnetosphere,
$\rho_{\rm eff}/B$ could be a small positive constant 
by the cross-field motion of charges (without pair creation);
in this case, the current is slightly sub-GJ with electron emission 
from the neutron star surface,
as MH04a,b suggested.
On the other hand, if $\Delta l_{\rm SG} > \rlc/40$ holds
in the outer magnetosphere,
$\rho_{\rm eff}/B$ takes a small negative value
by the discharge of the created pairs
(fig.~\ref{fig:C70three});
in this case, the current is super-GJ with ion emission from the surface,
as demonstrated by H06.
Since no studies have ever successfully constrained 
the gap transfield thickness,
there is room for further investigation on this issue.

\section*{Acknowledgments}
The author is grateful to Drs.
A.~K.~Harding, A.~N.~Timokhin, for fruitful discussion.
This work is supported by the Theoretical Institute for
Advanced Research in Astrophysics (TIARA) operated under Academia Sinica
and the National Science Council Excellence Projects program in Taiwan
administered through grant number NSC 94-2752-M-007-001.

%
%
%
%

\end{document}